\documentclass[epj]{svjourarxiv}
\usepackage{amsmath}
\usepackage{amssymb}
\usepackage{mathtext}
\usepackage{graphicx}

\begin{document}

\title{Formation of adsorbate structures induced by external electric field 
in plasma-condensate systems}
\author{Vasyl O.~Kharchenko$^{1,2}$\thanks{vasiliy@ipfcentr.sumy.ua},
Alina V.~Dvornichenko$^2$, Vadym N. Borysiuk$^{2}$ }
\institute{$^{1)}$Institute of Applied Physics, National Academy of Sciences of Ukraine, 
58 Petropavlivska St., 40000 Sumy, Ukraine\\
$^{2)}$Sumy State University, 2 Rimskii-Korsakov St., 40007 Sumy, Ukraine}
\authorrunning{V.O.Kharchenko, A.V.Dvornichenko, V.M.Borysiuk}
\titlerunning{Adsorbate structures formation in plasma-condensate systems}
\abstract{
We present a new model of plasma-condensate system, by taking into account an anisotropy 
of transference reactions of adatoms between neighbor layers of multi-layer system, caused by the strength 
of the electric field near substrate. We discuss an influence of the strength of the electric field 
onto first-order phase transitions and conditions for adsorbate patterning in plasma-condensate systems. 
It is shown that separated pyramidal-like multi-layer adsorbate islands can be formed in the plasma-condensate 
system if the strength of the electric field near substrate becomes larger tan the critical value, 
which depends on the interaction energy of adsorbate and adsorption coefficient.    
  }
 \date{}
\maketitle

\section{Introduction}

Nano-structured thin films attract increased attention in science and engineering due to 
their unique properties which make possible their usage in modern electronic devices 
\cite{1mph,2mph,3mph,4mph}. Experimental studies and analytical treatment together with 
numerical simulations show that by using different techniques for thin films growth one 
can observe patterning on the growing surface. These patterns can be either transient or 
stationary. It depends on the system control parameters, which are usually related to: 
flux of adatoms  toward growing surface; type of adsorbed particles (type of atoms/ions), 
that defines the activation energies of adsorption and desorption; temperature; etc. 
Moreover, in real experiments one always deals with multi-layer growth, where the 
minimization principle of the surface energy leads to an appearance of the vertical 
current of adatoms from top toward bottom layers. At the same time from the naive 
physical consideration one can state, that the concentration of adsorbate on the bottom 
layer is always larger then one on the top layer. This difference results in a bias in 
a standard vertical diffusion of adsorbate between layers. The competition of these two effects 
also affects the growing surface morphology \cite{Wolgraef2003,Wolgraef2004,SS15}. 

Most of experimental studies and theoretical modeling and predictions of nano-sized 
adsorbate islands formation were done for high-vacuum and low-vacuum vapor 
deposition in gas-condensate systems. Here by varying the pressure in a chamber and 
temperature one can get either ordered pattern of vacancies inside adsorbate matrix 
\cite{5mph}, or elongated adsorbate structures \cite{6mph,7mph} or 
nano-dots  \cite{8mph,9mph}. At the same time analytical studies and 
numerics show that real systems manifest self-organization processes with formation of 
stable complexes of adadoms in addition to equilibrium 
adsorption and desorption reactions \cite{TH96,11mph}. It was shown that non-equilibrium 
reactions only govern realization of stable surface patterns in numerical simulations 
of pattern formation in one-layer gas-condensate model 
\cite{BHKM97,HME98_1,HME98_2,Wolgraef2002,PhysScr2012}. In Refs.\cite{PRE12,SS14} authors 
discussed controlling of the type of surface patterns and linear size of structures by 
varying adsorption coefficient, interaction energy of adsorbate, non-equilibrium reactions 
rate and intensity of internal fluctuations. In multi-layer low-vacuum gas-condensate 
systems with vertical anisotropic current of adsorbate, caused by gas phase pressure 
large-sized adsorbate islands on the upper layer are formed on the adsorbate matrix 
with separated vacancy cluster on bottom layers \cite{SS15}.

In order to produce separated adsorbate islands with small lateral linear size the 
plasma-condensate devices are used. In the framework of this technique ions, sputtered 
by magnetron, attain growing surface, located in a hollow cathode \cite{Per8,Perekrestov}. 
The electric field near 
substrate leads to desorption of a part of adatoms back to plasma, their additional 
ionization, and adsorption on the high levels of multi-layer growing surface \cite{Per8}. 
Hence, the strength of the electric field controls the anisotropy strength of the 
vertical diffusion of adatoms between neighbor layers with preferential motion from 
lower layers to upper ones, in contrast to the case for gas phase pressure induced 
anisotropy in low-vacuum gas-condensate systems. Processes of surface patterning in 
plasma-condensate devices were considered mostly by experimental studying (see, for 
example, Refs.\cite{Perekrestov,Perekrestov2}). It was shown a possibility to get 
separated adsorbate islands with linear size around hundreds nanometers. In recent 
work \cite{NRL17} authors provided numerical study of adsorbate islands formation 
in multi-layer plasma-condensate system by varying anisotropy strength of vertical 
diffusion of adsorbate between neighbor layers, caused by the strength of the electric 
field near substrate. At the same time there is no any discussion about the range of 
the main control parameters responsible for ordering processes on the surface during 
deposition.

The main goal of this work is to define conditions corresponding to formation of 
stationary separated (adsorbate or vacancy) structures on the growing surface in 
plasma-condensate systems. To that end we will derive the accurate model for adequate 
description of adsorbate concentration evolution in multi-layer plasma-condensate 
system by taking into account a decreasing of adsorbate concentration with number of 
layer and anisotropy in vertical diffusion of adatoms between neighbor layers induced 
by the electric field near substrate. The main attention in the article is paid to 
study an influence of the strength of the electric field near substrate onto stability 
of surface structures.

The work is organized in the following manner. In the next section we construct the 
model for the adsorbate concentration evolution on the intermediate layer of 
multi-layer plasma-condensate system by deriving the functional dependence of the 
concentration of adsorbate on neighbor layers \emph{versus} one on the current 
layer and discuss a physical limitation of the model. In Section 3 we perform analysis 
of the homogeneous system and discuss first-order phase transitions. 
In Section 4 we study stability of homogeneous stationary states to inhomogeneous 
perturbations and define conditions for stationary patterns formation in the system. 
We conclude in the last Section.

\section{Model}

Evolution of adsorbate concentration on the selected layer $n$ of $N$-layers system can be 
described by the reaction diffusion model of the standard form:
\begin{equation}
\partial_tx_n=R(x_n)-\nabla \mathbf{J}_n(x_n;\nabla), \label{eq1}
\end{equation}
where $x_n\in[0,1]$ defines through the ratio between adsorbed particles (adatoms) and total number 
of atomic cites in the fixed area of the layer lattice; $t$ is the time variable. Reaction term
$R(x_n)$ defines quasi-chemical reactions between adsorbed particles. For the $N$-layer system 
one should take into account the following reactions: (i) adsorption, when particles attach the substrate 
and become adatoms; (ii) desorption, when adatoms can desorb back into plasma; (iii) isotropic vertical 
motion of adatoms between neighbor layers. Moreover, by considering plasma-condensate systems 
one should take into account additional anisotropic vertical diffusion, caused by the action 
of the electric field near substrate. This field will lead to preferential motion of adatoms 
from bottom layers to top ones. From the naive consideration adsorption processes on $n$-layer 
can be described by the term $f_a=k_a x_{n-1}(1-x_n)(1-x_{n+1})$, where $k_a$ is the adsorption 
coefficient, proportional to the pressure of the plasma (density of the plasma). We take into account 
that particle can become adatom if: (i) there is nonzero adsorbate concentration on the previous layer ($x_{n-1}$); 
(ii) there are free sites for adsorption on the current layer $1-x_{n}$; (iii) there is free space on the next 
layer ($1-x_{n+1}$). Substratum mediated desorption processes are described by the term $f_d=
-k_d x_n x_{n-1} (1-x_{n+1})$, where desorption coefficient $k_d=k_{d0}\exp(U_i(\mathbf{r})/T)$ 
defines through desorption rate of non-interacting particles $k_{d0}$ and interaction potential $U_n(\mathbf{r})$. 
Here we also admit that nonzero adsorbate concentration on the current layer ($x_n$) 
can be desorb from existence previous layer ($x_{n-1}$) back to plasma through free space on the next layer 
$(1-x_{n+1})$. For the isotropic transference reaction between layers, which define the standard vertical diffusion, 
we have the standard one-dimension expression $f_{t}=k_{t}(x_{n-1}+x_{n+1}-2x_n)$ with the vertical diffusion 
coefficient $k_{t}$. To define the reaction rate for the electrical field induced anisotropic transference 
between layers let us proceed in the following way. The electric field is characterized by the strength 
$\mathbf{E}=\nabla\phi$, where $\phi$ is the difference in electric potentials. Hence, the additional 
current of adatoms on the current $n$ layer caused by the electric field is defined as 
$\mathbf{J}_{||}^E=\left(\mathbf{E}ZeD_{0}/T\right)x_n$, where $Z$ is the
valence of ion and $e$ is the electron charge. By taking into account that 
this transference is possible only on free sites on the corresponding layers
for the reaction term, describing anisotropic transference, we get 
$f_u=u[x_{n-1}(1-x_n)-x_n(1-x_{n+1})]$, where the rate $u=|\mathbf{E}|Ze/T$, 
corresponds to the anisotropy strength defined through the strength of the electrical field 
near substrate. 

To define the diffusion term $\mathbf{J}_n$ in Eq.(\ref{eq1}) we will use the 
approach, proposed in Refs.\cite{BHKM97,HME98_1,HME98_2,ME94,HM96}. 
We assume that the total lateral diffusion flux is described by the 
ordinary diffusion $-D_{0}\nabla x_n$ and diffusion caused by the interactions 
between adatoms on $n$ layer. The last one is defined through the force $F_n=-\nabla U_n$ and 
induces speed $v_n=(D_0/T)F_n$. The corresponding flux $v_nx_n$ is possible 
onto free ($1-x_n$) sites, only. Hence, the lateral diffusion flux of adsorbate on 
$n$ layer reads:
\begin{equation}
 \mathbf{J}_{n}=-D_{0}\nabla x_n-(D_{0}/T)\mu(x_n)\nabla U_n,\label{eq4}
\end{equation}
where the mobility of adatoms $\mu(x_n)=x_n(1-x_n)$ is introduced.
The unknown adsorbate interaction potential $U_n$ can be defined in the framework 
of self-consistent approximation, which frequently used by studying pattern formation in 
reaction-diffusion systems \cite{SS15,Wolgraef2003,Wolgraef2004,BHKM97,HME98_1,HME98_2,Wolgraef2002,PhysScr2012,PRE12,SS14,NRL17,ME94,HM96,MW2005,CWM2002,M2010},
pyramidal structures formation in phase field modeling \cite{PhysScr11,EPJB13,Cmph14,EPJB15}, etc. 
In the framework of this procedure the interaction potential $U_n(r)$ can be represented in the form:
$U_n(r)=x_{n-1}(r)\left\{-\int u(r-r')x_{n}(r'){\rm d}r'\right\},$ where $-u({\bf r})$ is 
the binary attractive potential. For $u(x)$ we use the Gaussian profile 
$u(r)=2\epsilon(4\pi r_{0}^2)^{-1/2}\exp\left(-r^2/4r_{0}^2\right)$, where 
$\epsilon$ is the interaction energy and $r_0$ is the interaction radius. 
By assuming that the adsorbate concentration varies slow within the interaction 
radius we expand the integral 
$$\int
u_1(\mathbf{r}-\mathbf{r}')x_1(\mathbf{r}'){\rm d}\mathbf{r}'\simeq \int
u_1(\mathbf{r}-\mathbf{r}')\sum_n \frac{(\mathbf{r}-\mathbf{r}')^n}{n!}\nabla^n
x_1(\mathbf{r}){\rm d}\mathbf{r}'$$ and retain only three
non-vanishing terms. Finally, we get:
\begin{equation}
\int u(\mathbf{r}-\mathbf{r}')x(\mathbf{r}'){\rm
d}\mathbf{r}'\simeq\epsilon
x(\mathbf{r})+\epsilon(1+r_{0}^2\nabla^2)^2x(\mathbf{r}).
\end{equation}
Hence, the substratum mediated interaction potential $U_n$ on $n$ layer becomes 
the form \cite{SS15,NRL17}:
\begin{equation}
U_n(\mathbf{r})\simeq-\epsilon x_{n-1}(\mathbf{r})\left[
x_n(\mathbf{r})+\mathcal{L}_{SH}x_n(\mathbf{r})\right],
\end{equation}
where $\mathcal{L}_{SH}=(1+r_0^2\nabla^2)^2$ is the Swift-Hohenberg operator \cite{LSH} 
and $\nabla U_n$ in Eq.(\ref{eq4}) is defined as follows: $\nabla U_n=-\epsilon x_{n-1}(\mathbf{r})\nabla\left\{
x_n(\mathbf{r})+(1+r_0^2\nabla^2)^2x_n(\mathbf{r})\right\}$.

The main goal of this work is to define the conditions of pattern formation 
on a surface during adsorption in plasma-condensate system with anisotropic 
vertical diffusion of adatoms between layers. In Ref.\cite{SS15} for two- and three-layers 
gas-condensate system authors shows the corresponding phase diagram illustrating domains 
of main system parameters when the growing surface can be structured with formation of 
separated adsorbate islands or separated vacancy islands inside adsorbate matrix
on each layer. In this article we are aimed to define the system parameters domains 
responsible for pattern formation on any layer of $N$-layers plasma-condensate system.
To this end we will derive the functional dependencies of the adsorbate concentration on 
the previous ($n-1$) and the next ($n+1$) layers through the adsorbate concentration on 
the current layer $n$ by taking into account that the concentration of adsorbate on each next 
layer is less than one on the current layer. 

By considering the multi-layer system the adsorbate concentration on each layer 
can be defined according to the fraction $x_n=S_n/S_0$ where $S_0\propto R_0^2$ is the square of 
the substrate with linear size $R_0$ and $S_n\propto R_n^2$ is the coverage square of $n$ layer 
by adsorbate. According to the principle of 
minimization of surface energy the adsorbate concentration on each next layer is less than one on 
the current layer.
Hence, considering multi-layer  adsorbate island as a pyramidal structure, we assume, that 
the linear size of the multi-layer adsorbate structure on each layer decreases by the small value $d$ 
with the layer number growth, $R_n=R_0-nd$. In the framework of this description 
the parameter $d$ defines the terrace width of the multi-layer pyramidal structure of adsorbate. 
Hence, the adsorbate concentration on any
$n$ layer is defined as: $x_n\simeq\left(1-nd/R_0\right)^2$. Next by using simple algebra 
one finds that the concentration of adsorbate on nearest layers to $n$ one is defined 
through $x_{n}$ as: 
$x_{n-1}=\left(\sqrt{x_n}+1/2\beta_0\right)^2$, $x_{n+1}=(\sqrt{x_n}-1/2\beta_0)^2$ with 
$\beta_0=2d/R_0<1$. In this case 
for the isotropic vertical diffusion described by $f_t$ one gets $1/2\beta_0^2$. It means, that 
according to condition that the concentration of adsorbate decreases with number of the layer 
the standard vertical diffusion is characterized by bias which tends to zero only for the 
layer by layer growth. Parameter $\beta_0$ defines the number $N$ of layers in 
multilayer plasma-condensate system through the relation $x_{N+1}=0$, as $N=2/\beta_0-1$, 
with $x_N=\beta_0^2/4$. 

By introducing the dimensionless constants 
$\varepsilon\equiv\epsilon/T$, $\alpha\equiv k_{a}/k_d^0$, $k_t'\equiv k_{t}/k_d^0$,
$u'=u/k_d^0$ and dropping all primes the reaction term $R(x)$ in Eq.(\ref{eq1}) for $x=x_n$ becomes the form:
\begin{equation}
\begin{split}
 R(x)&=\alpha(1-x)\nu(x)-x\nu(x)e^{-2\varepsilon x(\sqrt{x}+\frac{1}{2}\beta_0)^2}\\ 
 &+u\beta_0\sqrt{x}(1-2x)+\frac{1}{4}\beta_0^2(u+2),
\end{split}
\end{equation}
where $\nu(x)=(\sqrt{x}+1/2\beta_0)^2\left[1-(\sqrt{x}-1/2\beta_0)^2\right]$.
Next it is more convenient to measure time in units $t'\equiv tk_d^0$, distance in units $r'=r/L_d$, and 
by introducing diffusion length $L_d=\sqrt{D_0/k_d^0}$ the evolution equation for the 
adsorbate concentration Eq.(\ref{eq1}) reads:
\begin{equation}
\partial_tx=R(x)+D_0\nabla\cdot
\left[ \nabla x-\varepsilon\gamma(x)\nabla\left\{x+\mathcal{L}_{SH}x\right\}\right]+\xi(t),
\label{ev2}
\end{equation}
where $\gamma(x)=\mu(x)(\sqrt{x}+1/2\beta_0)^2$ and $\xi(t)$ is the delta-correlated 
Gaussian zero-mean noise with intensity $\sigma^2$ proportional to the bath temperature. 

\begin{figure}[!ht]
\centering\includegraphics[width=0.9\columnwidth]{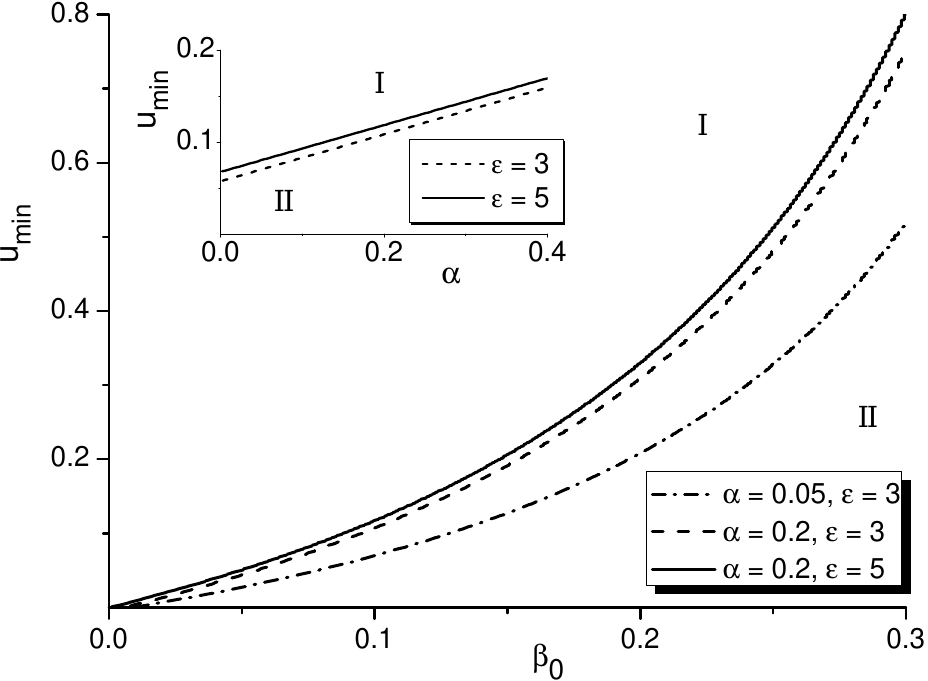}
\caption{Minimal value of the anisotropy strength $u_{min}(\beta_0)$: domain I 
corresponds to the system parameters when the concentration of adsorbate on the current  
layer does not exceed value $x_{max}$. 
Insertion shows the dependencies $u_{min}(\alpha)$ at $\beta_0=0.1$.}
\label{fig1}
\end{figure}

The derived model for evolution of the adsorbate concentration on the intermediate layer 
of multilayer system has one physical limitation, related to the limit value of the adsorbate 
concentration on the previous ($n-1$) layer, $x_{n-1}=1$, corresponding to the substrate. 
In such a case one gets the limit value 
for the adsorbate concentration on the current (first) layer, $x_{max}=(1-1/2\beta_0)^2$. 
This limit value allows one to define the minimal value of the anisotropy strength $u$ or 
the maximal value of the parameter $\beta_0$, when $x\leq x_{max}$. 
In Fig. \ref{fig1} we plot the dependence of the minimal value of the anisotropy strength 
$u_{min}$ on the parameter $\beta_0$. 
Here the dependence $u_{min}(\beta_0)$ bounds the domain  I of system parameters, 
when the concentration of adsorbate on the current layer does not exceed value $x_{max}$. 
From Fig. \ref{fig1} it is seen that even infinitely small value of $\beta_0$, which defines 
the difference in the adsorbate concentration on neighbor layers, requires non-zero  
value of the anisotropy strength $u_{min}$ for holding the relation $x_{n-1}\leq1$. At 
$\beta_0\to0$ the adsorbate concentration on any layer remains constant, defined  
by adsorption coefficient $\alpha$ and interaction strength $\varepsilon$. In such a case equation 
(\ref{ev2}) describes evolution of adsorbate concentration in one-layer-like model with 
$f_t=f_u/u=0$. It follows, that an increase in ether adsorption 
coefficient $\alpha$ or interaction strength $\varepsilon$ decreases the domain I corresponding to 
to multilayer growth. In the insertion 
in Fig. \ref{fig1} we show dependencies $u_{min}(\alpha)$ at fixed $\beta_0=0.1$. It is seen, 
that in the case of small pressure ($\alpha\to0$) $N$-layer growth is possible at non-zero 
anisotropy strength $u$ only (electric field induced transference of adatoms between neighbor 
layers).             

\section{Stability of the stationary homogeneous states}

\begin{figure}[!ht]
\centering\includegraphics[width=0.9\columnwidth]{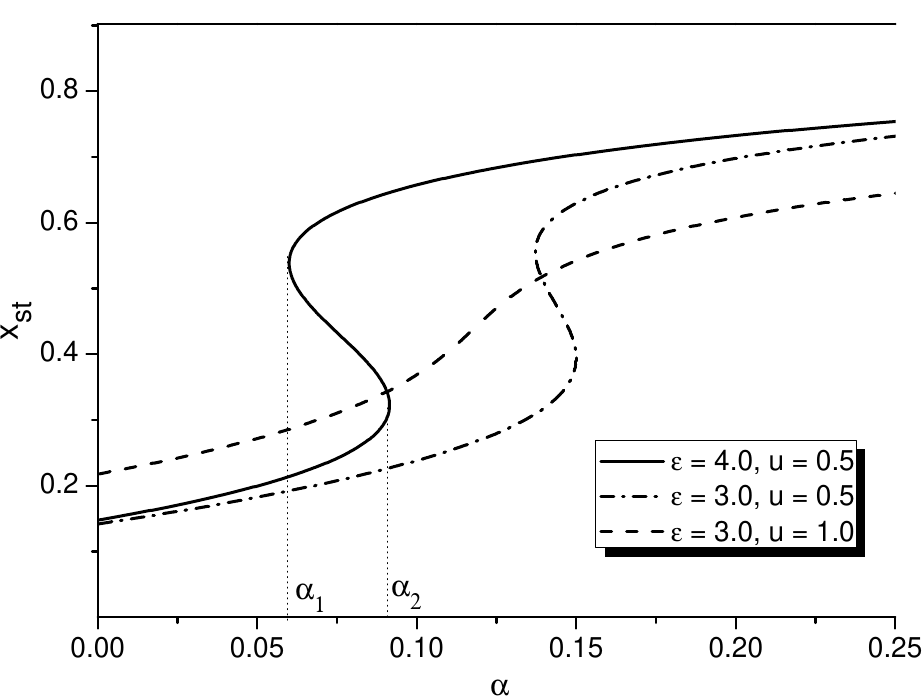}
\caption{Dependencies of the stationary adsorbate concentration on the intermediate layer $x_{st}$ 
on adsorption coefficient $\alpha$.}
\label{fig2}
\end{figure}

\begin{figure}[!ht]
\centering\includegraphics[width=\columnwidth]{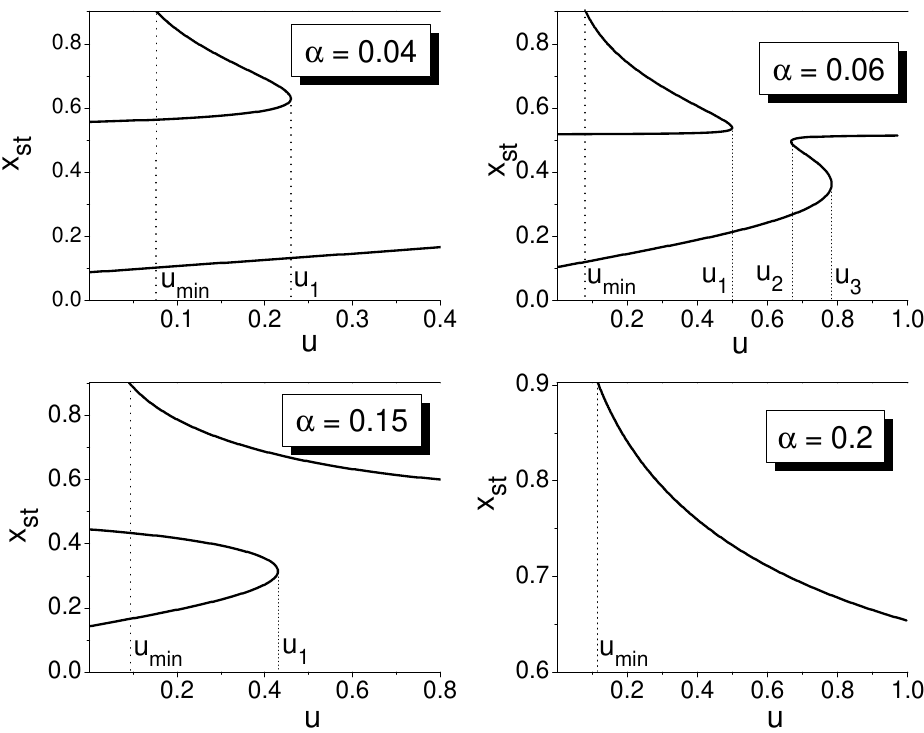}
\caption{Dependencies of the stationary adsorbate concentration on the intermediate layer $x_{st}$ 
\emph{versus} anisotropy strength $u$ at $\varepsilon=4.0$, $\beta_0=0.1$ and different values of 
the adsorption coefficient $\alpha$.}
\label{fig3}
\end{figure}

It is known that gas-condensate systems undergo  first-order phase transitions (see, for example, 
Refs.\cite{Wolgraef2002,Wolgraef2003,Wolgraef2004,PhysScr2012,PRE12,SS14}). In this section we will study 
homogeneous system and discuss dependencies of the stationary adsorbate concentration on control parameters 
and phase transitions in plasma-condensate system, described by Eq.(\ref{ev2}). 
To proceed we assume $\nabla\cdot\mathbf{J}=0$ 
and consider stationary limit with $\partial_tx=0$, leading to the equation $R(x)=0$. Typical 
dependencies of the stationary adsorbate concentration on the intermediate layer $x_{st}$ 
\emph{versus} adsorption coefficient $\alpha$ 
are shown in Fig. \ref{fig2} for different values of the anisotropy strength $u$.
It follows, that at small values of the anisotropy strength (see solid curve for $u=0.1$) the plasma-condensate 
system undergoes first-order phase transition with increase in the adsorption coefficient $\alpha$ from 
low-density state (plasma) to high-density state (adsorbate) at $\alpha=\alpha_1$. With a decrease in the 
adsorption coefficient one has transition from high- to low-density state at $\alpha=\alpha_2$. 
Hence if the adsorption coefficient takes values inside the interval $(\alpha_1,\alpha_2)$, then the system 
becomes bi-stable.  
A decrease in the interaction strength $\varepsilon$ requires larger values for adsorption coefficient 
$\alpha$ when the studied system is bi-stable; the width of the interval $(\alpha_1,\alpha_2)$ becomes smaller 
(compare solid and dash-dot curves in Fig.\ref{fig2}). An increase in the anisotropy strength $u$ leads 
to a degeneration of the interval for adsorption coefficient values, responsible for bi-stability of 
the system (compare dash-dot and dash curves in Fig.\ref{fig2}). Hence, at large $u$ system is aways 
mono-stable. In Fig. \ref{fig3} we present dependencies of the stationary adsorbate concentration $x_{st}$ 
\emph{versus} anisotropy strength $u$ at different values of adsorption coefficient $\alpha$. 
Here values of the homogeneous stationary adsorbate concentration are limited by the value 
$x_{max}$ which corresponds to the minimal value of the interaction strength 
$u_{min}(\alpha,\varepsilon,\beta_0)$. From Fig. \ref{fig3} it follows that 
at small values of adsorption coefficient with increase in the anisotropy strength one gets 
first-order phase transition from high-density state toward low-density state at $u=u_1$ (see panel 
for $\alpha=0.04$).  
An increase in adsorption coefficient results in more complicated picture of phase transformations 
(see panel for $\alpha=0.06$ in Fig. \ref{fig3}). Here at small $u$ one has the similar to the previous case 
picture with increased value $u_1$. For the case $u>u_1$ one gets additional first-order 
phase transition from low- to high-density state and the width of the hysteresis loop $(u_2,u_3)$ 
depends on both adsorption coefficient $\alpha$ and interaction energy $\varepsilon$. This effect 
is caused by the competition between adsorption and anisotropic transference. Further growth in $\alpha$ 
leads to transformation of the dependence $x_{st}(u)$ and an increase in anisotropy strength 
leads to first-order phase transition from low-density state to high-density state at $u=u_1$
(see panel for $\alpha=0.15$). At large values of $\alpha$ explicit decreasing dependence $x_{st}(u)$ 
is realized in the high-density phase. 
 
\begin{figure}[!ht]
\centering\includegraphics[width=0.9\columnwidth]{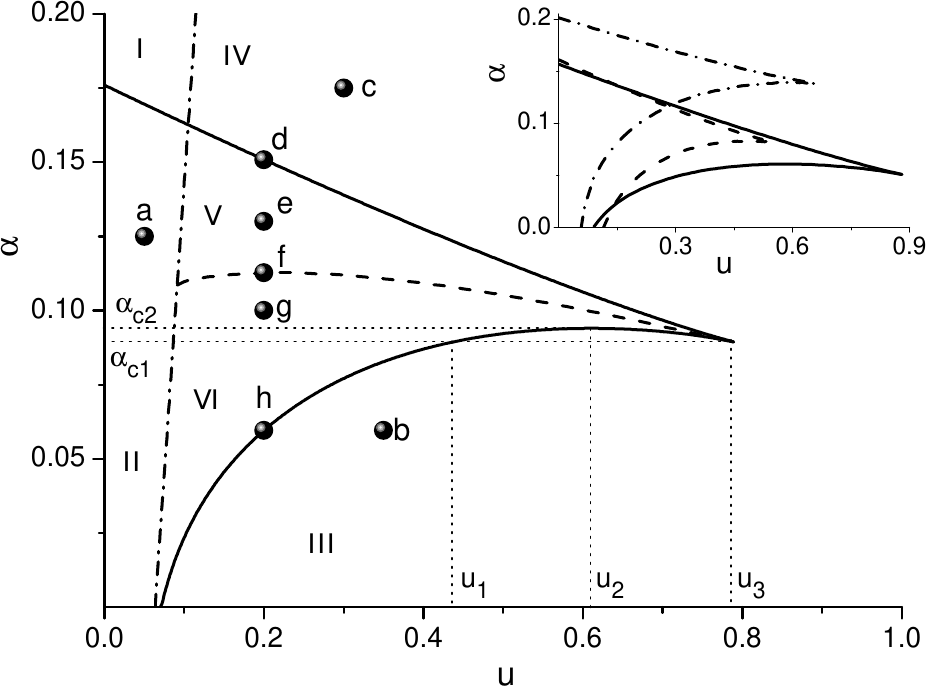}
\caption{Phase diagram $\alpha(u)$ for the plasma-condensate system obtained at $\varepsilon=3.5$ and 
$\beta_0=0.1$. In insertion the diagram $\alpha(u)$ is shown at: $\varepsilon=4$, $\beta_0=0.1$ (solid 
lines); $\varepsilon=3$, $\beta_0=0.1$ (dash-dot lines); $\varepsilon=3$, $\beta_0=0.15$ (dash lines).}
\label{fig4}
\end{figure}

Obtained dependencies $x_{st}(\alpha)$ and $x_{st}(u)$ allow one to calculate the phase diagram, 
illustrating domains of system parameters, where the studied plasma-condensate system is bi-stable. 
In Fig. \ref{fig4} we plot such a diagram in plane $(u,\alpha)$ at fixed values of the adsorbate interaction 
energy $\varepsilon$ and the difference in adsorbate concentrations on
neighbor layers $\beta_0$.  
It follows that the whole plane is divided onto six different domains by four 
curves. Here solid lines (binodals) define the system parameters ($\alpha_1(u)$ and $\alpha_2(u)$) when the 
studied systems becomes bi-stable (domains V and VI). Dash line corresponds to the spinodal. 
Dash-dot curve relates to the dependence $u_{min}(\alpha)$ (see insertion in Fig. \ref{fig1}).  
It  bounds the bi-stability domains 
from the left by values of the system parameters when $x\leq x_{max}$. 
\begin{figure}[!ht]
\centering\includegraphics[width=0.9\columnwidth]{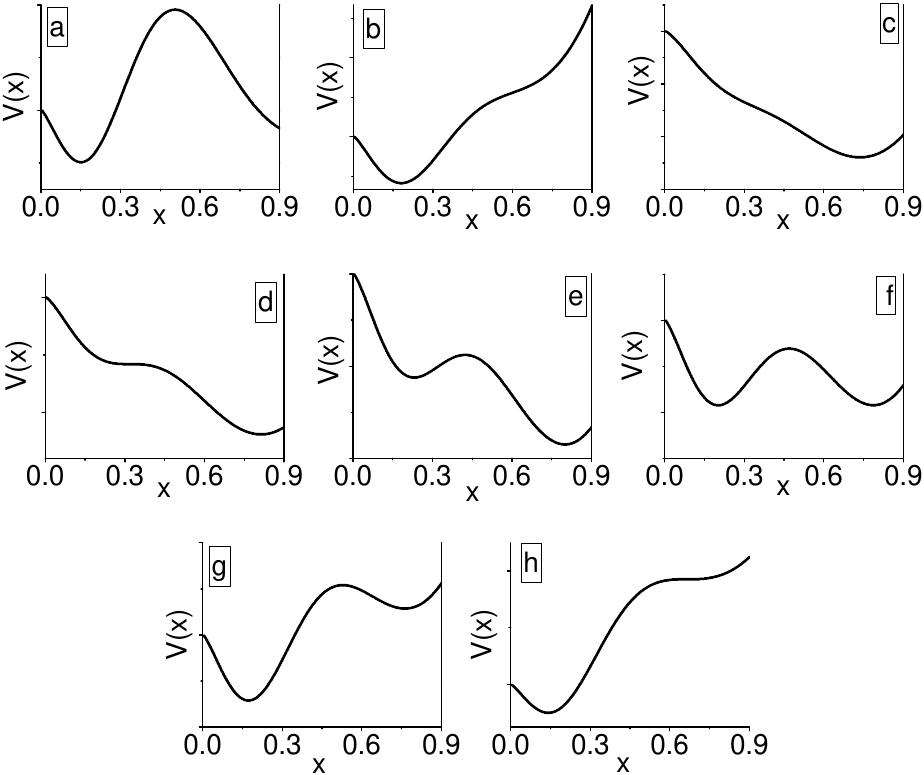}
\caption{Effective potentials $V(x)$ at $\varepsilon=3.5$, $\beta_0=0.1$ and different values of 
$\alpha$ and $u$.}
\label{fig5}
\end{figure}
From Fig. \ref{fig4} it follows that inside the domain I the relation 
$x_{st}\leq x_{max}$ does not hold and no physically realized stationary value of adsorbate 
concentration does exist. The domains of system bi-stability V and VI are bounded from the 
right by the value $u_3$. By analyzing the bi-stability domain VI one can 
define the characteristic values of the adsorption coefficient: $\alpha_{c1}=\alpha(u_{1})=\alpha(u_{3})$ 
and $\alpha_{c2}=\alpha(u_{2})$, where $u_{2}$ can be defined from the equation $\partial_u\alpha=0$.
At fixed $\alpha$ taken from interval $\alpha_{c1}<\alpha<\alpha_{c2}$ an increase in 
the anisotropy strength $u$ will lead to re-entrant picture of first-order phase transitions 
(see the dependence $x_{st}(u)$ at $\alpha=0.06$ in Fig. \ref{fig3}).

For detailed analysis of system states we calculate the effective potential $V(x)=-\int R(x)$ in each 
domain and at values $\alpha$ and $u$, corresponding to binodals and spinodal. 
Minimums of $V(x)$ in Fig. \ref{fig5} define stable states. 
In domain II the system is characterized by one stable and one unstable states (one maximum and one 
minimum of the dependence $V(x)$ in Fig. \ref{fig5}a). Domains III and IV correspond to the system parameters
when the system is characterized by the single stationary low- and high-density state, accordingly
(see panels b and c in Fig. \ref{fig5}). 
By fixing the anisotropy strength $u<u_3$ with decrease in the adsorption coefficient $\alpha$ 
we move from domain IV of single high-density state (point c) to bistability domain V (point e) through 
the binodal (point d). In this case the additional minimum of the potential $V(x)$ appears and 
system is in high-density state (the minimum at large $x$ is deeper). 
The two minimums of the potential $V(x)$ becomes equivalent at 
values of the system parameters from the spinodal (dash curve)  (see panel $f$ in Fig. \ref{fig5}). 
Further decrease in the adsorption coefficient leads to transition of the system toward low-density state 
(see panel $g$ in Fig. \ref{fig5}).   
 
In the insertion in Fig.\ref{fig4} we show a change in the bi-stability domain by varying 
interaction strength $\varepsilon$ and adsorbate difference on neighbor layers $\beta_0$. 
Here solid curves correspond to $\varepsilon=4$, $\beta_0=0.1$; dash-dot lines relate to 
$\varepsilon=3$, $\beta_0=0.1$ and dash ones are obtained at  $\varepsilon=3$, $\beta_0=0.15$.
It follows, that with an increase in $\beta_0$ the bi-stability domain shrinks 
in both $\alpha$ and $u$ (compare dash-dot and dash curves). An increase in the 
interaction strength leads to extension of the bi-stability domain in values of anisotropy 
strength, shrink in adsorption coefficient and appearance of the interval $(\alpha_{c1},\alpha_{c2})$ 
of re-entrant first-order phase transitions (see Fig. \ref{fig3} at $\alpha=0.06$).    

\section{Stability analysis and pattern formation}

\begin{figure}[!ht]
\begin{center}
a)\includegraphics[width=0.9\columnwidth]{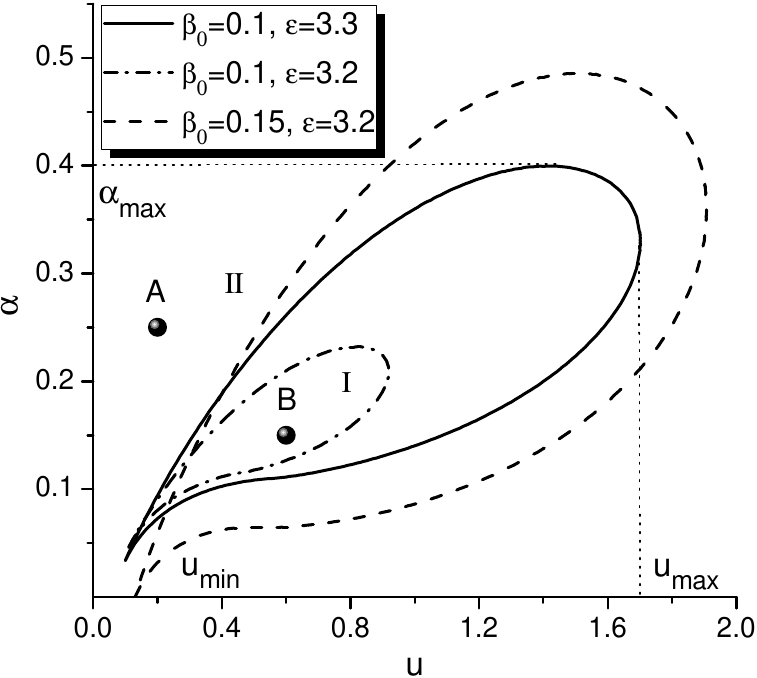}\\
b)\includegraphics[width=0.45\columnwidth]{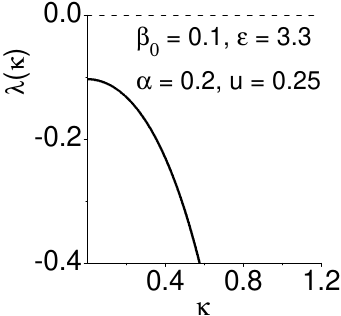}
c)\includegraphics[width=0.45\columnwidth]{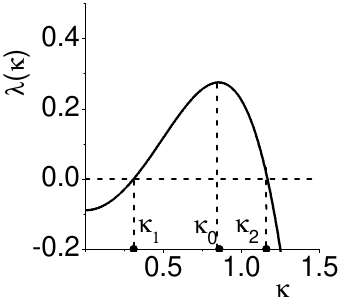}
\end{center}
\caption{Phase diagram for the plasma-condensate system illustrating domain of system 
parameters relevant to formation of stationary separated adsorbate structures (domain I). 
In domain II during exposing time adsorbate will homogeneously cover the substrate. 
Dependencies of the stability exponent obtained in points $A$ and $B$ are shown in panels 
(b) and (c) respectively.}
\label{fig6}
\end{figure}

To define the conditions for adsorbate structures formation during condensation in studied plasma-condensate 
system in this section we will provide the linear stability analysis of homogeneous stationary states $x_{st}$ 
to in-homogeneous perturbations. To that end in the framework of the standard linear stability analysis  we will 
consider the deviation of the adsorbate concentration from the stationary value: 
$\delta x=x-x_{st}\propto e^{\lambda(k)t}e^{ikr}$, where $k$ is the wave number, $\lambda(k)$ is the stability exponent. 
Next, by expanding the reaction term $R(x)$ in the vicinity of the stationary state $x_{st}$ with $R(x_{st})=0$ from 
the evolution equation for the deviation $\delta x$, obtained from Eq.(\ref{ev2}) one gets the stability exponent 
$\lambda(k)$ in the form:
\begin{equation}
\begin{split}
\lambda(\kappa)&={\rm d}_xR(x)|_{x=x_{st}}\\
  &-D_0\kappa^2\left[1-\varepsilon\gamma(x_{st})\left\{1+(1-\rho_0^2\kappa^2)^2\right\}\right],
\label{lambda}
\end{split}
\end{equation}
where $\rho_0=r_0/L_d$\footnote{For most of metals and semi-conductors the interaction radius 
$r_0\sim10^{-9}$m, and the diffusion length $L_d\sim10^{-7}-10^{-6}$m.} 
and $\kappa=kL_d$. The stability exponent $\lambda(\kappa)$ can be used to 
characterize the stationary morphology of the surface during condensation processes. If 
$\lambda(\kappa)<0$ for all wave numbers $\kappa$ then any spatial perturbation degenerates in time 
and no spatial instabilities are possible in the stationary regime. Here any adsorbate structures, 
which can be formed at initial stages of the system evolution disappear at large time scales and 
adsorbate will cover the substrate homogeneously. In the case $\lambda(\kappa)>0$ at $\kappa\in(0,\kappa_c)$ 
one gets picture typical for phase separation. For the case $\lambda(\kappa)>0$ at 
$\kappa\in(\kappa_1,\kappa_2)$ one can expect formation of separated structures on the surface 
with the period of spatial modulations (mean distance between structures) $\kappa_0$ defined 
from ${\rm d}_\kappa\lambda(\kappa)=0$. By taking into account that $\rho_0^4\to0$ the analysis 
of Eq.(\ref{lambda}) shows, that in the limit of large wave numbers $\kappa$ the stability 
exponent $\lambda(\kappa)$ is negative.  Hence, to ensure the conditions of pattern formation 
in the studied plasma-condensate system one needs: (i) ${\rm d}_xR(x)|_{x=x_{st}}<0$, which defines 
that the corresponding stationary state $x_{st}$ is stable; (ii) $\lambda(\kappa_0)>0$. These 
two conditions allow one to compute the phase diagram, which define domain of  
control parameters values when during adsorption processes in plasma-condensate system one gets stationary 
separated adsorbate structures or vacancy clusters inside adsorbate matrix. 
The corresponding phase diagram in a plane $(u,\alpha)$ is shown 
in Fig. \ref{fig6}a for different values of interaction energy $\varepsilon$ and the coefficient $\beta_0$. 
Here for the system parameters taken from the bounded domain I stationary structures will be form  
on the growing surface at the condensation in multi-layer plasma-condensate system. In the domain II 
no patterns can be observed. The corresponding dependencies of the stability 
exponent $\lambda(\kappa)$ obtained in points $A$ and $B$ are shown in Figs. \ref{fig6}b and \ref{fig6}c, 
respectively. From Fig. \ref{fig6}b it is seen, that in the domain II the stability exponent takes 
negative values in the whole interval of the re-normalized wave numbers $\kappa$. 
In the domain I the dependence $\lambda(\kappa)$ crosses 
zero value twice and takes positive values inside the bounded interval of the wave numbers 
($\kappa_1$,$\kappa_2$). The mean period of themstationary structures location 
(mean distance between separated structures) is defined by the value $r_0\propto1/\kappa_0$ in the ${\bf r}$-space. 

\begin{figure*}[!t]
\begin{center}
a)\includegraphics[width=0.3\textwidth]{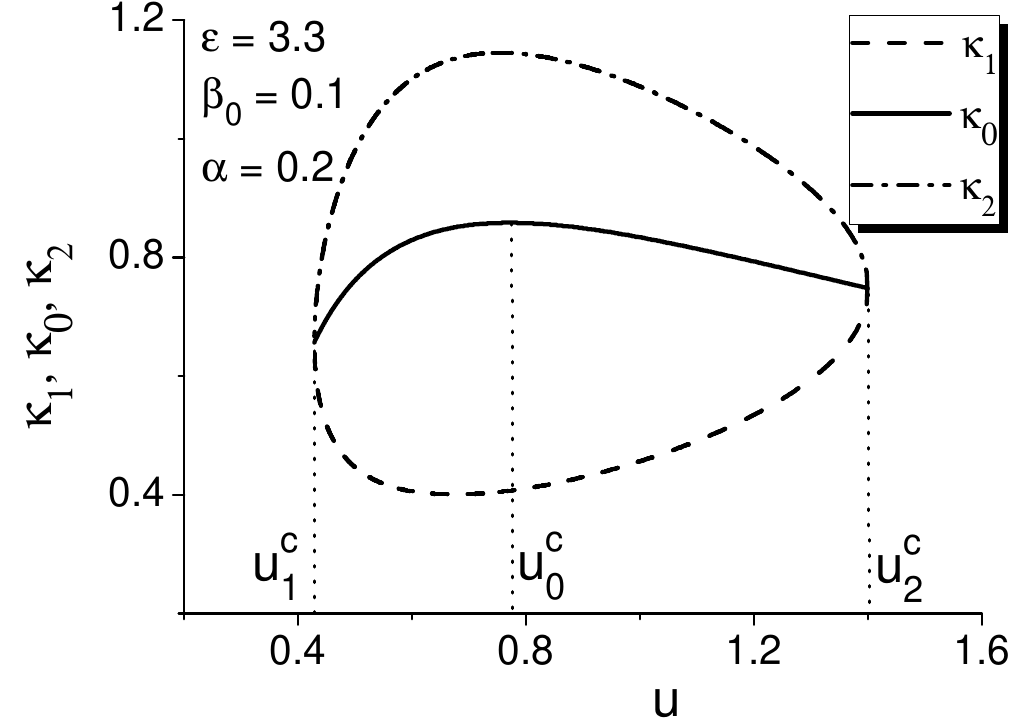}
b)\includegraphics[width=0.3\textwidth]{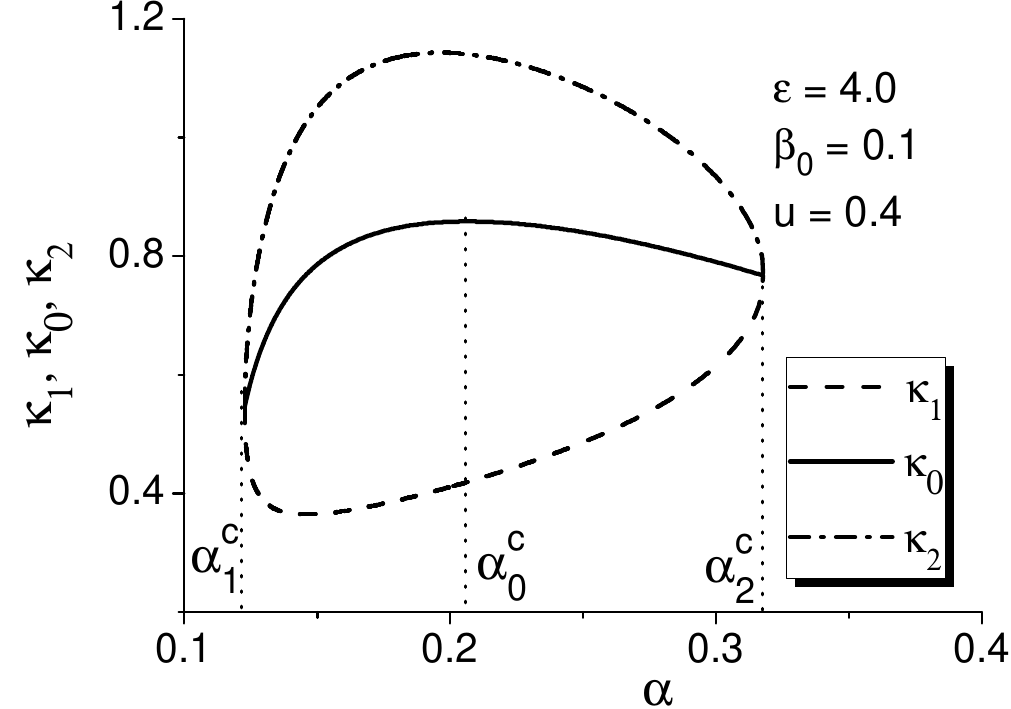}
c)\includegraphics[width=0.3\textwidth]{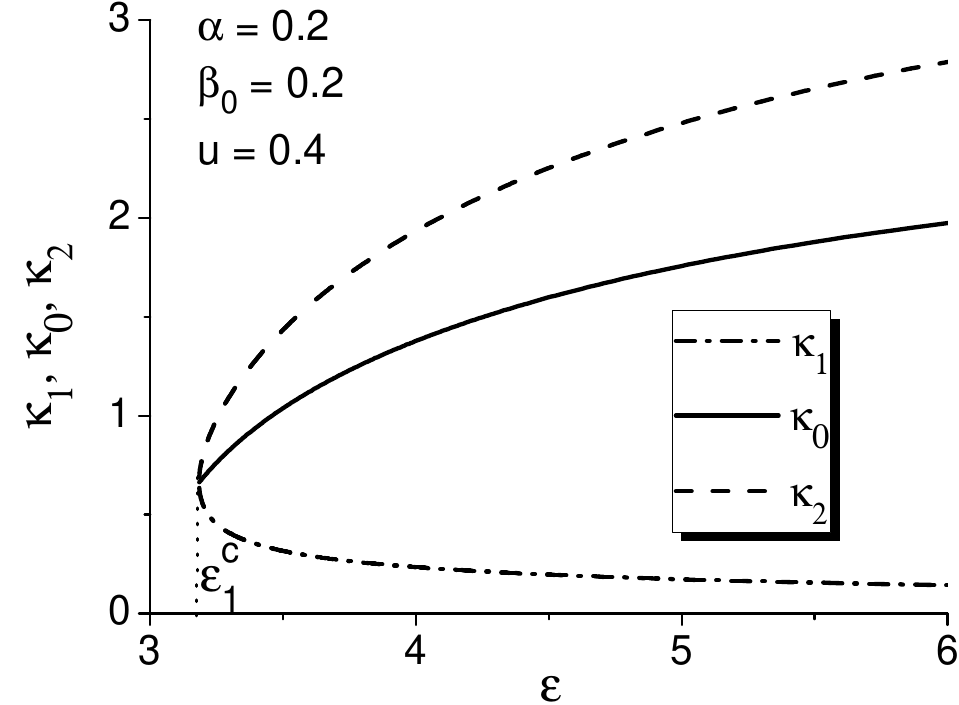}
\end{center}
\caption{Dependencies of the wave numbers $\kappa_1$ and $\kappa_2$, which define the interval 
of the positive values of the stability exponent $\lambda(\kappa)$, and $\kappa_0$, corresponds to 
the mean period of adsorbate structures location, \emph{versus}: (a) anisotropy strength $u$; 
(b) adsorption coefficient $\alpha$; (c) adsorbate interaction energy $\varepsilon$.}
\label{fig7}
\end{figure*}
\begin{figure*}[!t]
\begin{center}
\includegraphics[width=0.9\textwidth]{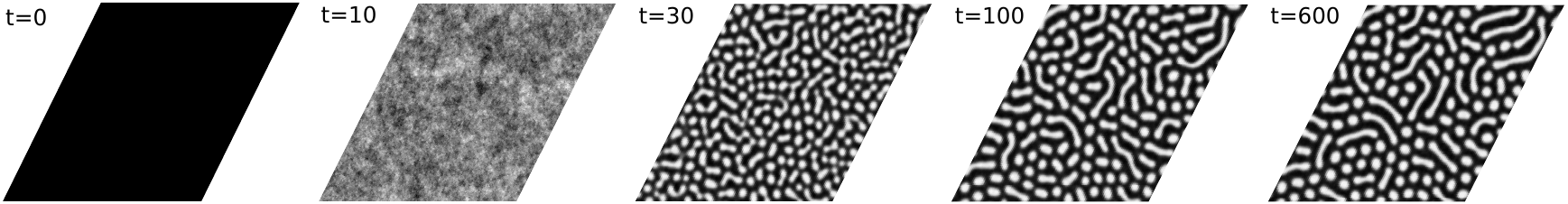}
\end{center}
\caption{Numerical simulations of the evolution of the morphology of the intermediate layer of multi-layer 
system at $\alpha=0.2$, $\varepsilon=3.5$, $\beta=0.1$, $u=1$ and $\sigma^2=0.01$.}
\label{fig8}
\end{figure*}

One should mentioned, that in the case of isotropic transference reactions 
(at $u=0$) no stationary spatial instability can be possible for any values of adsorption coefficient 
$\alpha$, interaction energy $\varepsilon$ and coefficient $\beta_0$. Let us discuss dependence 
$\alpha(u)$ in Fig. \ref{fig6} at $\beta_0=0.1$ and $\varepsilon=3.3$ in details (see solid curve). 
The domain of pattern formation 
(domain I) is bounded by the minimal value of the adsorption coefficient $\alpha(u_{min})$ 
and the maximal values $\alpha_{max}$ and $u_{max}$. 
Hence, by fixing the anisotropy strength $u$ (fixing the strength of the electric field near 
substrate) and by increasing the adsorption coefficient we get the 
following transformations in the surface morphology. At low values of $\alpha$ (domain II 
under the solid curve) there is no 
enough adsorbate concentration on the layer to induce pattern formation and adsorbate with small 
concentration covers the layer homogeneously. With increase in the adsorption coefficient 
inside the domain I one gets separated adsorbate island on the current layer, meaning formation 
of pyramidal-like multi-layer adsorbate islands. At $\alpha>\alpha_{max}$ the large pressure 
produces large adsorbate concentration and neither desorption no interaction no anisotropic 
transference can stabilize adsorbate structures, formed at initial stages of the system evolution 
and adsorbate covers whole the layer, meaning growing of the surface in layer-by-layer scenario. 
Hence, an increase in adsorption coefficient leads to re-entrant patterning in the multi-layer 
plasma-condensate system. The same effect can be seen by varying the strength of the electric field, 
described by $u$, at fixed values of adsorption coefficient $\alpha$: small anisotropy can 
not induce pattern formation processes; at $u>u_{max}$ the strong anisotropy in vertical diffusion leads 
to a decrease in the adsorbate concentration and we get process analogous to the case with small $\alpha$. 
From obtained results it follows, that a decrease in both the adsorbate interaction energy and the 
difference in the adsorbate concentration on neighbor layers leads to shrink in the size of the domain 
of pattern formation.    

Next, we will consider dependencies of the wave numbers $\kappa_1$ and $\kappa_2$, bounding domain of 
positive values of the stability exponent $\lambda(\kappa)$ and the value $\kappa_0$, defining the mean 
distance between structures, shown in Fig. \ref{fig7}. 
From Fig. \ref{fig7}a it follows, that at $u<u_{1}^c$ the whole dependence $\lambda(\kappa)$ lies under  the line 
$\lambda=0$. At $u>u_{1}^c$ the stability exponent becomes positive in the interval $(\kappa_1,\kappa_2)$, 
which extends with growth in $u$. Here $\kappa_0$ increases with $u$ meaning formation of new adsorbate  islands 
on the layer, that leads to a decrease in the mean distance between them in ${\bf r}$-space. 
At $u=u_{0}^c$, which is defined by the condition ${\rm d}_u\kappa_0=0$ the interval for the 
positive values of the stability exponent becomes shrink and the distance between adsorbate clusters 
falls down, meaning formation of smaller number of adsorbate islands with larger period of their location 
due to anisotropy in transference reactions between layers. For the case $u>u_{2}^c$ the stability exponent 
becomes negative and strong anisotropy leads to adsorbate structures disappear in the stationary regime.
The dependence of the wave numbers $\kappa_1$,  $\kappa_2$ and $\kappa_0$ on the adsorption coefficient 
$\alpha$ shown in Fig. \ref{fig7}b is similar to the previous case. A decrease dependence 
$\kappa_0(\alpha)$ here means an increase of the linear size of adsorbate islands.
From the dependence of wave numbers on the interaction energy $\varepsilon$ (see Fig. \ref{fig7}c) 
it follows, that if the condition $\varepsilon>\varepsilon_1^c$ holds then the stability 
exponent becomes positive similar to the case 
shown in Fig. \ref{fig7}a. An increase in $\varepsilon$ extends the interval $(\kappa_1,\kappa_2)$ and leads to 
growth in $\kappa_0$, meaning formation of large number of adsorbate structures with small averaged 
distance between them in ${\bf r}$-space.  

One should stress that the provided analysis of the stability of the homogeneous stationary state to 
inhomogeneous perturbation (see Fig. \ref{fig6}) allows one to determine conditions 
(values of the control parameters), when spatial instability will lead to formation of stationary 
multi-layer structures in the plasma-condensate system. At the same time this analysis does not 
give any information about the morphology of the layer defined by the type of structures, namely, either 
separated adsorbate structures on the layer, either separated holes in adsorbate matrix, 
or elongated percolating islands (typical for phase separation picture). 

To illustrate the process of pattern formation during condensation in plasma-condensate 
system we perform numerical simulation. To this end we solve Eq.(\ref{ev2}) numerically on two-dimension 
grid with size $L\times L$ with $L=256$ sites, triangular symmetry (to provide formation of spherical structures) 
and periodic boundary conditions. We provide numerical integration with time-step $\Delta t=10^{-3}$ and 
mesh size $\Delta x=0.5$ by using relation $L_d/r_0=40$, that gives the relation $L=12.8L_d$. 
We assume, that at time instant $t=0$ the concentration of adsorbate 
on the intermediate layer equals zero. Snapshots of the system evolution at 
$\alpha=0.2$, $\varepsilon=3.5$, $\beta=0.1$, $u=1$ and $\sigma^2=0.01$ 
(inside domain I in Fig. \ref{fig6}a) are shown in Fig. \ref{fig8}.
Here with the grey color we show the concentration of adsorbate counting from minimal (black) 
to maximal (white) color for each time instant. It follows, that at initial times the adsorbate 
starts to cover the current layer (see panel at $t=10$). 
When the concentration of adsorbate attends supersaturation the small separated adsorbate clusters start 
to appear (see panel at $t=30$). These clusters grow in time and 
rearrange (see panel at $t=100$). At large time scales one gets the stationary picture of adsorbate clusters 
distribution on the current layer (see panel at $t=600$), meaning formation of pyramidal-like 
islands of adsorbate during condensation in multi-layer plasma-condensate system. From numerical 
simulations we get the stationary mean adsorbate concentration,  
$\langle x_{st}\rangle\simeq0.5943$, that gives the number of the current layer 
$n=5$ with $N=19$ is the total number of layers in adsorbate multi-layer 
structures. By using $L_d\simeq5\cdot10^{-7}m$ we get the mean radius of stationary 
spherical structures $\langle R_n\rangle\simeq0.18$ $\mu m$, the mean radius of 
adsorbate structures on the first layer $\langle R_0\rangle\simeq0.24$ $\mu m$ and the terrace width 
for the adsorbate structures $d\simeq0.012$ $\mu m$.

\section{Conclusions}

In this article we present a new model of plasma-condensate system, by taking into account an anisotropy 
of transference reactions of adatoms between neighbor layers of multi-layer system, caused by the strength 
of the electric field near substrate. 
The derived 
model was used to describe pattern formation 
processes on an intermediate layer of multi-layer system. We focused our attention on detailed study 
of an influence of the anisotropy strength on the morphology change of the growing surface. 
We have discussed first-order phase transitions 
plasma-condensate and define the phase diagram illustrating the bounded domain of main system parameters 
when separated adsorbate cluster will be formed on the current layer of multi-layer system meaning 
growth of pyramidal-like adsorbate islands. It is shown that an increase in both adsorption coefficient 
proportional to the plasma pressure and strength of the anisotropy in transference of adatoms between neighbor 
layers, caused by the strength of the electric field near substrate, leads to re-entrant picture of 
pattern formation on the surface during adsorption/desor\-ption processes in plasma-condensate systems.
It is shown, that multi-layer nano-sized pyramidal-like adsorbate structures can be formed if the  
strength of the electric field near substrate exceeds the critical value. 

\section*{Acknowledgments}

Support of this research by the Ministry of Education and Science of Ukraine, 
project No. 0117U003927, is gratefully acknowledged.


\end{document}